\documentclass[12pt]{article}

\usepackage{epsfig}
\usepackage{amsmath}
\usepackage{amssymb}
\usepackage{graphicx}
\newcommand\npb[3]{{\it Nucl. Phys. }{\bf B #1} (#2) #3}

\newcommand\prd[3]{{\it Phys. Rev. }{\bf D #1} (#2) #3}

\def\permil{\%\raise.10ex\hbox{$_{\scriptstyle 0}$}}
\def\be{\begin{equation}}
\def\ee{\end{equation}}
\def\bea{\begin{eqnarray}}
\def\eea{\end{eqnarray}}
\def\bsp{\begin{split}}
\def\esp{\end{split}}

\newcommand\epjc[3]{{\it Eur. Phys. J.}{\bf C #1} (#2) #3}  
     
\begin{document}
\titlepage
\begin{flushright}
DESY-09-043 \\
March 2009
\end{flushright}
\vspace{-.5cm}
\begin{center}
{\Large \bf Saturation and linear transport equation.}\\
\vspace*{0.5cm}
 K.\ Kutak \\
 {\it DESY Notkestrasse 85, 22607 Hamburg, Germany}\\

\end{center}
\vspace*{1cm}
\centerline{(\today)}

\vskip1cm
\begin{abstract}
\noindent
We show that the GBW saturation model provides an exact solution to the one-dimensional linear transport equation. We also show that 
it is motivated
by the BK equation considered in the saturated regime when the diffusion and the splitting term in the diffusive 
approximation are balanced by the nonlinear term.

\end{abstract}
\section{Introduction}
Perturbative Quantum Chromodynamics (pQCD) at high energies can be formulated in
coordinate space in the dipole picture \cite{Mueller}. If we in particular focus on Deep Inelastic Scattering the scattering process
can be described in this picture as interaction of virtual photon which has just enough energy to dissociate into a 'color
dipole' with the hadronic target carrying most of the total energy. The interaction process is described here by the  
dipole-nucleus scattering amplitude. 
The formal formulation and application of the dipole picture leads to high energy pQCD evolution equations for the dipole amplitude and in particular to a 
BFKL equation for this quantity \cite{BFKL} which together with a certain factorisation theorem allows to calculate observables in the dipole picture.
However, leaving out the formal approach one can also model the dipole amplitude. 
The Golec-Biernat W\"usthoff saturation model developed roughly 10 years ego \cite{GBW} is a quite well tested model for the dipole amplitude  which 
includes saturation effects. It was motivated by requirements that at the high energy limit of QCD the total cross section for hadronic processes
should obey unitarity requirements. At present there are much more sophisticated approaches to introduce these requirements
in a description of scatterings at high energies \cite{JIMWLK,Bal,Kov,AGL,BartKut}. However, one can still ask the question if there is any dynamics 
behind the GBW model or to put it differently is there any equation to which formula proposed by Golec-Biernat and W\"usthoff is a solution? And what is the role of the initial conditions? In this article we want to answer these questions.\\
The Letter is organized as follows: in the next section we show that the GBW model provides exact 
solution to the one-dimensional transport equation. We show this for the unintegrated gluon density and for the dipole amplitude in the momentum space. 
In the third section we relate the derived transport equation to the BK 
equation which allows us to conclude that the GBW model is well based in pQCD.
\section{GBW model and a transport equation}
\subsection{Momentum space analysis}

The GBW amplitude following form GBW cross section and related to it by $\sigma(x,r)=2\int d^2b N(x,r,b)$ reads (here we are interested in the original 
formulation without  evolution in the
hard scale \cite{BKGB}):
\be
N(x,r,b)=\theta(b_0-b)\Bigg[1-\exp\left(-\frac{r^2}{4R_0^2}\right)\Bigg]
\label{eq:GBW98}
\ee  
where $b$ is the impact parameter of the collision defined as distance between center of the proton with radius $b_0$ and center of a dipole 
scattering on it, 
$r$ is a transversal size of the dipole, $x$ is the Bjorken variable, 
$R_0(x)\!=\!\frac{1}{Q_0}\left(\frac{x}{x_0}\right)^{\lambda/2}$ 
is the so called saturation radius and its inverse defines saturation scale, $Q_s(x)\!=\!1/R_0(x)$ and $x_0$, $\lambda$ are free parameters. 
This amplitude saturates for large dipoles $r\!\gg\!2R_0$ and exhibits geometrical scaling which has been confirmed by data \cite{GBKS}.
\subsubsection{Transport equation for unintegrated gluon density}
The dipole amplitude (\ref{eq:GBW98}) can be related to the unintegrated gluon density  which convoluted with the $k_T$ dependent 
off-shell matrix elements allows to calculate observables in the high energy limit of QCD. This relation is the following (after assumption that the dipole is 
much smaller than the target) \cite{Braun,KKM01}:
\be
f(x,k^2,b)=\frac{N_c}{4\alpha_s\pi^2}k^4\nabla^2_k\int \frac{d^2{\bold r}}{2\pi}\exp(-i{\bold k}\cdot{\bold r})\frac{N(x,r,b)}{r^2}
\label{eq:ugd}
\ee
where $r$ and $k$ are two-dimensional vectors in transversal plane of the collision and $r\equiv|\bold r|$, $k\equiv|\bold k|$\\ 
Performing this transformation we obtain the known result \cite{GBF}:
\be
f(x,k^2,b)=\frac{N_c}{2\pi^2\alpha_s}\theta(b_0-b)R_0^2(x)k^4\exp\left[-R_0^2(x)k^2\right]
\label{eq:ugd2}
\ee
 Now motivated by the fact that this formula exhibits a maximum both as a function of $x$ for fixed $k^2$ and as a function
of $k^2$ for fixed $x$, we differentiate $f(x,k^2,b)$ with respect to $x$ and $f(x,k^2,b)/k^2$ with respect to $k^2$.
We obtain:
\be
\partial_x f(x,k^2,b)=\frac{\lambda f(x,k^2,b)(1-R_0^2(x)k^2)}{x Q_0^2}
\label{eq:poch1}
\ee
\be
\partial_{k^2} \frac{f(x,k^2,b)}{k^2}=\frac{f(x,k^2,b)(1-R_0^2(x)k^2)}{k^4Q_0^2}
\label{eq:poch2}
\ee
Dividing eqn. (\ref{eq:poch1}) by (\ref{eq:poch2}) and rearranging the terms and defining ${\cal F}(x,k^2,b)=f(x,k^2,b)/k^2$, $Y=\ln x_0/x$, $L=\ln k^2/Q_0^2$
we obtain:
\be
\partial_Y {\cal F}(Y,L,b)+\lambda\partial_L {\cal F}(Y,L,b)=0
\label{eq:trans}
\ee
which is the first order linear wave equation also known as the transport equation. As it is linear it cannot generate saturation dynamically but it can propagate well the initial condition leading to a successful phenomenology \cite{GBW}. 
It describes the change (wave) in the particle distribution flowing into and out of the phase space volume with velocity $\lambda$. 
This wave propagates in one direction. The quantity ${\cal F}(x,k^2,b)$ gains here the interpretation of a number density of gluons with momentum 
fraction $x$ with the transversal momentum $k^2$ at distance $b$ from the center of the proton. 
The general solution of (\ref{eq:trans}) can be found by the method of characteristics and is given by:
\be
{\cal F}(Y,L,b)={\cal F}_0(L-\lambda Y,b)
\label{eq:sol}
\ee
One can go back from (\ref{eq:trans}) to (\ref{eq:ugd2}) using following initial condition at $x=x_0$: 
\be
{\cal F}(x\!=\!x_0,k^2,b)=
\frac{N_c}{2\pi^2\alpha_s}\theta(b_0-b)k^2\exp(-k^2)\\
\label{eq:gbwini}
\ee
This initial condition has saturation built in, since the gluon density vanishes for small $k^2$.
One can also try a different initial condition at large $k^2$:
\be
{\cal F}(x=x_0,k^2,b)\sim \frac{1}{k^2}\\
\label{eq:gbwini2}
\ee
which gives solution without the saturation effect:
\be
{\cal F}(x,k^2,b)\sim \frac{1}{k^2}\left(\frac{x}{x_0}\right)^{\lambda}\\
\label{eq:gbwsol2}
\ee 
Knowing the properties of the linear first order partial differential equation we see that the property of saturation
of GBW was a consequence of the wave solution which relates $x$ and $k^2$ supplemented by initial conditions with saturation built in. 
We also see that the {\it critical line} of the GBW saturation model visualizing, the dependence of the saturation scale on $x$, 
$Q_s(x)=Q_0\left(\frac{x_0}{x}\right)^{\lambda/2}$ is in fact from the mathematical point of view the characteristics of the 
transport equation.
\subsubsection{Transport equation for the dipole amplitude in momentum space}
Similar investigations can be repeated for the momentum space representation of the dipole amplitude  $N(x,r,b)$ 
which we denote by $\phi(x,k^2,b)$. 
\be
\phi(x,k^2,b)=\int\frac{d^2{\bold r}}{2\pi}\exp(-i{\bold k}\cdot{\bold r})\frac{N(x,r,b)}{r^2}
\label{eq:ugd}
\ee
A nonlinear pQCD evolution equation like the Balitksy-Kovchegov (BK) equation written for $\phi$ (in large target approximation) takes quite simple form and 
can be related directly to the statistical formulation of the high energy limit of QCD (see \cite{SM2} and references therein). 
Applying this transformation to (\ref{eq:GBW98}) we obtain:
\be
\phi(x,k^2,b)=\frac{1}{2}\theta(b_0-b)\,\Gamma\left[0,\frac{k^2}{Q_0^2}\left(\frac{x_0}{x}\right)^\lambda\right]
\label{eq:dipms}
\ee
where $\Gamma(0,z)=\int_z^{\infty}\frac{dt}{t}e^{-z}$.
To obtain the partial differential equation we proceed similarly as before and upon differentiating (\ref{eq:dipms}) with 
respect to $k^2$ and also with respect to $x$ and introducing $Y=\ln x_0/x$, $L=\ln k^2/k_0^2$ we obtain:
\be
\partial_Y\phi(Y,L,b)+\lambda \partial_L \phi(Y,L,b)=0
\ee
which is, as before, the transport equation which supplemented with the initial condition, 
\be
\phi(x\!=\!x_0,k^2,b)=\frac{1}{2}\theta(b_0-b) \Gamma\left[0,\frac{k^2}{Q_0^2}\right]
\ee
gives back (\ref{eq:dipms}). 
\subsection{Relation to pQCD}
It is tempting to investigate the relation between  (\ref{eq:trans}) and the high energy pQCD evolution equations  like \cite{Bal,Kov,BartKut}. 
Let us focus here in particular on the form of the BK equation in 
large cylindrical target approximation for the dipole amplitude in momentum space for which the nonlinear term is just a simple local quadratic expression.
The BK equation for the dipole amplitude in the momentum space reads:
\be
\partial_Y\phi(Y,k^2,b)=\overline\alpha\chi\left(-\frac{\partial}{\partial_{\log k^2}}\right)\phi(Y,k^2,b)-\overline\alpha\phi^2(Y,k^2,b)
\label{eq:BKdip}
\ee
where $\overline\alpha=\frac{N_c\alpha_s}{2\pi}$  and $\chi(\gamma)=2\psi(1)-\psi(\gamma)-\psi(1-\gamma)$ is the characteristic function of the BFKL 
kernel which allows for emission of dipoles and therefore drives the rise of the amplitude. The role of the nonlinear term is 
roughly to allow for multiple scatterings of dipoles which contributes with negative sign and slows down the rise of the amplitude. This equation 
provides unitarization of the dipole amplitude \cite{BKphen} for fixed impact parameter and admits traveling wave solution in the diffusion approximation\cite{SM1}.
\subsubsection{Analytic approach}
The analytic solution of (\ref{eq:BKdip}) within the diffusion approximation relying on expanding the kernel of (\ref{eq:BKdip}) up to second 
order and mapping it to the Fisher-Kolmogorov equation has been obtained by Munier and Peschanski \cite{MP3}. It reads:
\be
\phi(Y,k^2,b)=\theta(b_0-b)\sqrt{\frac{2}{\overline\alpha\chi''(\gamma_c)}}
\ln\left(\frac{k^2}{Q_s^2(Y)}\right) 
\left(\frac{k^2}{Q_s^2(Y)}\right)^{\gamma_c-1}
\exp\left[-\frac{1}{2\overline\alpha\chi''(\gamma_c)Y}\ln^2\left(\frac{k^2}{Q_s^2(Y)}\right) 
\right]
\label{eq:solBK1}
\ee
where $\gamma_c=0.373$ and $Q_s^2(Y)$ is emergent
saturation scale given by:
\be
Q_s^2(Y)=Q_0^2e^{-\bar\alpha\chi'(\gamma_c) Y-\frac{3}{2\gamma_c}\log Y
-\frac{3}{(1-\gamma_c)^2}
\sqrt{\frac{2\pi}{\bar\alpha\chi^{\prime\prime}(\gamma_c)}}\frac{1}{\sqrt{Y}}
+{\cal O}(1/Y)}\ 
\label{eq:satscal}
\ee
By inspection we see that (\ref{eq:solBK1}) does not obey the transport equation. The problem is caused by the diffusion term.  
However, we can consider the asymptotic regime called "front interior"  \cite{MP3,ebert}, region where transverse momenta $k$ is close to the saturation scale 
$Q_s(Y)$ and rapidity $Y$ is large and where the condition $\ln^2\left(\frac{k^2}{Q_s^2(Y)}\right)/2\overline\alpha\chi''(\gamma_c)Y\!<\!\!\!<\!1$ is satisfied.
In this regime (\ref{eq:solBK1}) simplifies to:
\be
\phi(Y,k^2,b)=\theta(b_0-b)\sqrt{\frac{2}{\overline\alpha\chi''(\gamma_c)}}\left(\frac{k^2}{Q_s^2(Y)}\right)^{\gamma_c-1}
\log [k^2/{Q_s^2(Y)}]
\label{eq:solutionBK2}
\ee
Proceeding as in the previous sections we obtain the following wave equation:
\be
\partial_Y {\phi}(Y,L,b)+\lambda_{BK}\partial_L {\phi}(Y,L,b)=0
\label{eq:transBK}
\ee 
 where $\lambda_{BK}=\partial \log Q_s^2(Y)/\partial Y$. In the limit where $\lambda_{BK}$ does not depends on energy \cite{IJM} we obtain:
\be
\lambda_{BK}=-\overline\alpha\chi'(\gamma_c)
\label{eq:asymlimit}
\ee
\subsubsection{Numerical results}
\begin{figure}[t!]
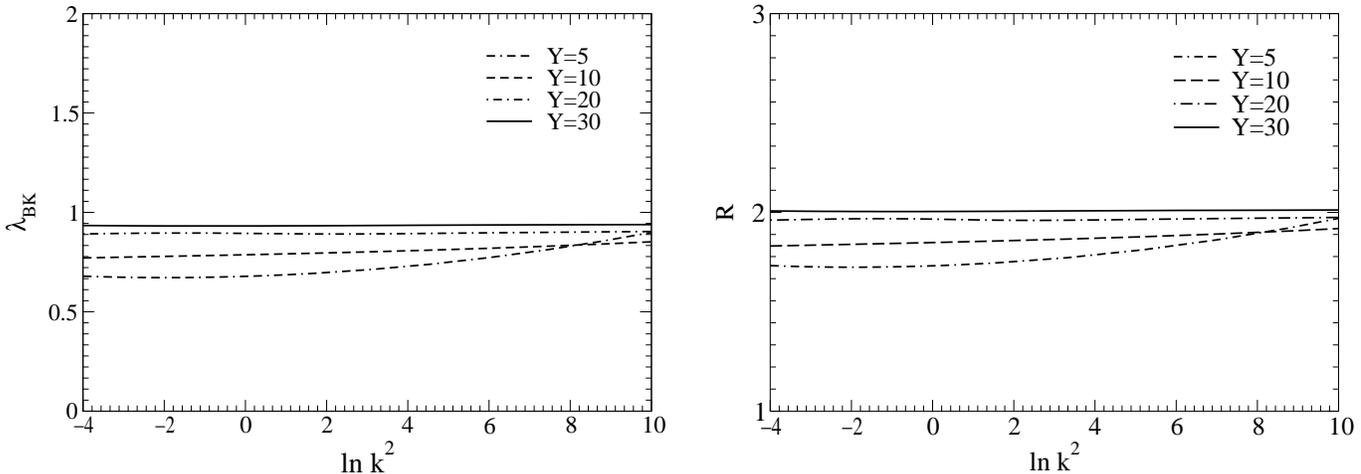

  \begin{picture}(30,30)
    \put(-50, 30){
      \includegraphics{velocitiesc.eps}
    }
     
   \put(210, 30){
      \includegraphics{ratioc.eps}
    }   
      \end{picture}
\vspace{6cm}
\caption{\em \small Velocity of wave plotted as a function of $\ln k^2$ for four different values of rapidity: Y=5, 10, 20, 30 
(a). Ratio $R$ defined in the text for four different values of rapidity: Y=5, 10, 20, 30. This plot shows
 that at saturation region the BK equation reduces to transport equation. }
\label{fig:plotvel}
\end{figure}

Now let us investigate the exact numerical solution of the BK equation in the diffusion approximation to see how its terms arrange to reduce to the transport equation in the 
saturation regime.
Following \cite{SM1} we represent the BFKL kernel of the BK equation as a power series around $\gamma_c$:
\be
\chi(-\partial_L)\phi(Y,L,b)=\left[\chi(\gamma_c)+(-\partial_L-\gamma_c)\chi'(\gamma_c)+\frac{1}{2!}(-\partial_L-\gamma_c)^2\chi''(\gamma_c)+...\right]\phi(Y,L,b)
\label{eq:BK}
\ee 
Taking terms up to the second order and rearranging them we obtain the following equation:
\be
-\frac{\partial_Y{\phi}(Y,L,b)}{{\partial_L{\phi}(Y,L,b)}}=\overline\alpha\chi_0'(\gamma_c)-\overline\alpha\frac{D\phi(Y,L,b)-
\phi^2(Y,L,b)}{\partial_L{\phi}(Y,L,b)}
\label{eq:BKdiff222}
\ee 
where $D=\chi_0(\gamma_c)-\gamma_c\chi_0'(\gamma_c)+\frac{1}{2}(-\partial_L-\gamma_c)^2\chi_0''(\gamma_c)$.\\
The right hand side of this equation using the analogy to the transport equation should define velocity of the wave. 
Solving numerically (\ref{eq:BKdiff222}) with the initial condition $\phi(0,L,b)=\theta(b_0-b)e^{-L^2}$ we obtain
results for wave velocities which are shown on Fig. (\ref{fig:plotvel}a).\\
Defining ratio:
\be
R\!\equiv\!\frac{D\phi(Y,L,b)- \phi^2(Y,L,b)}{\chi_0'(\gamma_c)\partial_L{\phi}(Y,L,b)}
\ee
and plotting it we observe  (Fig. (\ref{fig:plotvel}b)) 
that in the saturation region (small $L$, large $Y$) the second term in (\ref{eq:BKdiff222}) is proportional to the first one and approaches 
$2\overline\alpha\chi_0'(\gamma_c)$ which together with the first term gives the velocity of a wave traveling towards higher values of $\log k^2$. 
Using this fact we can write in that in asymptotic limit $Y\!\to\!\infty$:
\be
-\frac{\partial_Y{\phi}(Y,L,b)}{\partial_L{\phi}(Y,L,b)}=-\overline\alpha\chi_0'(\gamma_c)\!\equiv\!\lambda_{BK}
\label{eq:BKdiff2}
\ee 
which is the same as (\ref{eq:transBK}).
The numerical value {\small$\lambda_{BK}=0.92$} is clearly different as compared to GBW approach where velocity {\small$\lambda=0.27$} 
(four flavor fit) is a free parameter which is to
be determined by data. 
However, in the case of the BK equation {\small$\lambda$} is a number only in the asymptotic region. When contributions beyond asymptotic ones are taken  
into account like diffusion, proper kinematics in emission of dipoles, renormalization group effects \cite{enberg} it becomes a quite complicated 
function of energy. 
\section{Conclusions}
In this  note we have shown that the GBW saturation model is the exact solution of a one-dimensional linear transport equation of the form (\ref{eq:trans}).
We conclude that since (\ref{eq:trans}) is a linear equation the saturation property has to be provided in the initial condition.
We found that for the GBW model this equation is universal for the unintegrated gluon density $f(x,k^2,b)$ and the dipole amplitude in momentum space 
$\phi(x,k^2,b)$ but the details of the shape of the wave depends on the initial condition which is different
for each of them. 
We also studied the relation of the transport 
equation to the BK equation in the diffusion approximation. We have shown that in the 
region of phase space where diffusion and splitting processes  are of the same  order as the nonlinear term, the GBW model is consistent with the BK equation. 
\section{Acknowledgments}
I would like to thank Krzysztof Golec-Biernat for useful comments, suggestions and for careful reading of manuscript.
Discussions and comments by  G\"osta Gustafson and Hannes Jung are also acknowledged. I would like also to thank Anna Ochab-Marcinek and 
Rikard Enberg for useful correspondence.
\begin{small}

\end{small}
\end{document}